# Comment: Gibbs Sampling, Exponential Families and Orthogonal Polynomials

**Patrizia Berti, Guido Consonni, Luca Pratelli and Pietro Rigo**

## 1. GENERAL REMARKS

Let $K$ be a reversible Markov kernel on a measurable space $(S, \mathcal{B})$ with stationary distribution $P$. Regard $K$ as a linear operator, $K : L_2(P) \to L_2(P)$, and suppose that $L_2(P)$ admits an orthonormal basis of (real) eigenfunctions $\varphi_0, \varphi_1, \ldots$ for $K$. Thus, $\varphi_0 = 1$ and

$$K\varphi_j(s) = \int \varphi_j(t) K(s, dt) = \beta_j \varphi_j(s),$$
$$s \in S, j = 1, 2, \ldots,$$

for some (real) eigenvalue $\beta_j$. Under mild additional conditions,

(1) $\quad 4\|K^\ell(s, \cdot) - P\|^2 \leq \sum_{j>0} \beta_j^{2\ell} \varphi_j^2(s) \quad$ for all $s \in S$,

where $\|\cdot\|$ is total variation norm and $K^\ell$ the $\ell$th iterate of $K$. Using (1) is quite natural in MCMC where information on the convergence rate is crucial. For the 2-component Gibbs sampler, however, one drawback is that $K$ is generally not reversible.

Diaconis, Khare and Saloff Coste (DKS, in the sequel) go through this problem by noting that the marginal chains (the $x$-chain and the $\theta$-chain) are reversible, and bounding the marginal chains yields bounds on the bivariate chain. More importantly, in a few examples, DKS are able to diagonalize the marginal kernels, that is to evaluate their eigenvalues and eigenfunctions. A basic fact is that, in such examples, the eigenfunctions agree with the orthogonal polynomials corresponding to the marginals of $P$.

*Patrizia Berti is Professor, Dipartimento di Matematica Pura ed Applicata "G. Vitali", Universita' di Modena e Reggio-Emilia, via Campi 213/B, 41100 Modena, Italy e-mail: patrizia.berti@unimore.it. Guido Consonni is Professor, Dipartimento di Economia Politica e Metodi Quantitativi, Universita' di Pavia, via S. Felice 5, 27100 Pavia, Italy e-mail: guido.consonni@unipv.it. Luca Pratelli is Professor, Accademia Navale, viale Italia 72, 57100 Livorno, Italy e-mail: pratel@mail.dm.unipi.it. Pietro Rigo is Professor, Dipartimento di Economia Politica e Metodi Quantitativi, Universita' di Pavia, via S. Felice 5, 27100 Pavia, Italy e-mail: prigo@eco.unipv.it.*

Following this route, DKS give explicit sharp estimates, both lower and upper, on the convergence rate of a 2-component Gibbs sampler. Their results are interesting, elegant and promising of some generalizations. On the other hand, since an explicit diagonalization is required, they cover a few particular cases only. In real problems, when sampling from $P$, the available information is usually not enough for a diagonalization. Moreover, it is not clear how to handle the $k$-component Gibbs sampler for $k > 2$ using DKS's argument. Thus, in addition to DKS's bounds, it could be useful to have other estimates of the convergence rate, possibly less sharp but with a broader scope.

Here, we adopt the latter point of view and look for estimates based on classical drift conditions. In a sense, we investigate the extent of DKS's words in Section 1: "Finding useful $V$ and $q$ is currently a matter of art" (where $V$ and $q$ are the ingredients of a drift condition). We will play the devil's advocate, of course.

## 2. PLAIN ERGODICITY

As far as possible, our notation agrees with DKS's. Thus, $(\mathcal{X}, \mathcal{F})$ and $(\Theta, \mathcal{G})$ are measurable spaces, with $\mathcal{F}$ and $\mathcal{G}$ countably generated, and $P$ is a probability measure on the product $\sigma$-field $\mathcal{F} \otimes \mathcal{G}$. We let

$$X : \mathcal{X} \times \Theta \to \mathcal{X} \quad \text{and} \quad T : \mathcal{X} \times \Theta \to \Theta$$

denote the canonical projections. It is assumed that $P$ has a density $f$ with respect to $\mu \times \pi$, where $\mu$ is a $\sigma$-finite measure on $\mathcal{F}$ and $\pi = P \circ T^{-1}$ is the prior. Also, $m(x) = \int f(x, \theta) \pi(d\theta)$ is the density of $P \circ X^{-1}$ with respect to $\mu$. As DKS, we assume $0 < m(x) < \infty$ for all $x \in \mathcal{X}$.

We always refer to the Gibbs sampler with kernel

$$J((x,\theta),C)$$
$$= \frac{1}{m(x)} \int \int I_C(a,b) f(x,b) f(a,b) \mu(da) \pi(db)$$

where $(x,\theta) \in \mathcal{X} \times \Theta$ and $C \in \mathcal{F} \otimes \mathcal{G}$. Loosely speaking, this is the version of the Gibbs sampler where the initial state $(x,\theta)$ is first updated into $(x,b)$ and then into $(a,b)$. Abusing notation, since $J$ only depends on $x$, we write $J(x,\cdot)$ instead of $J((x,\theta),\cdot)$. Note that DKS denote our $J$ by $\widetilde{K}$.

A first point to be settled, before discussing rates of convergence, is ergodicity. Indeed, for Gibbs sampling to make sense, $J$ should be *ergodic*, in the sense that

$$\|J^\ell(x,\cdot) - P\| \to 0 \quad \text{for all } x \in \mathcal{X} \text{ as } \ell \to \infty.$$

A simple equivalent condition is in Berti, Pratelli and Rigo (2008, Theorem 4.5). Letting $\mathcal{N} = \{C \in \mathcal{F} \otimes \mathcal{G} : P(C) = 0\}$, $J$ is ergodic if and only if

(2) $$\overline{\sigma(X)} \cap \overline{\sigma(T)} = \mathcal{N}$$

where $\overline{\sigma(X)} = \sigma(\sigma(X) \cup \mathcal{N})$ and $\overline{\sigma(T)} = \sigma(\sigma(T) \cup \mathcal{N})$. A more transparent version of (2) is

$$P(X \in A) = 0 \quad \text{or} \quad P(T \in B) = 0$$
$$\text{whenever } A \in \mathcal{F}, B \in \mathcal{G} \text{ and}$$
$$P(A \times B) = P(A^c \times B^c) = 0.$$

Moreover, a working sufficient condition for (2) is

(3) $$\{X \in A\} \cap \{T \in B\} \subset \{f > 0\}$$
$$\subset \{X \in A\} \cup \{T \in B\}$$

for some $A \in \mathcal{F}$, $B \in \mathcal{G}$ with $P(A \times B) > 0$; see Berti, Pratelli and Rigo (2008), Corollary 3.7.

## 3. UNIFORM ERGODICITY

Let $K$ be a Markov kernel on $(S, \mathcal{B})$ with stationary distribution $P$. If $K(s,\cdot) \geq \epsilon Q(\cdot)$, $s \in S$, for some $\epsilon > 0$ and probability $Q$ on $\mathcal{B}$, then $\|K^\ell(s,\cdot) - P\| \leq (1-\epsilon)^\ell$, $s \in S$. Coming back to the Gibbs sampler, this fact implies:

PROPOSITION 1. *If $m$ is bounded, then*

$$\|J^\ell(x,\cdot) - P\| \leq \left(1 - \frac{u}{\sup m}\right)^\ell \quad \text{for all } x \in \mathcal{X}$$
$$\text{where } u = \sup_{B \in \mathcal{G}} \pi(B) \inf_{x \in \mathcal{X}, \theta \in B} f(x,\theta).$$

PROOF. This is essentially Remark 4.6 of Berti, Pratelli and Rigo (2008). For definiteness, we repeat the calculations here. Let $(S,\mathcal{B}) = (\mathcal{X} \times \Theta, \mathcal{F} \otimes \mathcal{G})$, $K = J$ and $u(B) = \pi(B) \inf_{\mathcal{X} \times B} f$. It can be assumed $u(B) > 0$ for some $B \in \mathcal{G}$ (otherwise, $u = 0$ and the Proposition 1 holds trivially). Fix one such $B$ and define $\epsilon = u(B)/\sup m$ and $Q(\cdot) = P(\cdot \mid T \in B)$. Then,

$$J(x,C) \geq J(x, C \cap \{T \in B\})$$
$$= \frac{1}{m(x)} \int \int I_C(a,b) I_B(b) f(x,b) f(a,b)$$
$$\cdot \mu(da) \pi(db)$$
$$\geq \frac{\inf_{\mathcal{X} \times B} f}{\sup m} P(C \cap \{T \in B\}) = \epsilon Q(C)$$

for all $x \in \mathcal{X}$ and $C \in \mathcal{F} \otimes \mathcal{G}$. Since $P$ is stationary for $J$, it follows that

$$\|J^\ell(x,\cdot) - P\| \leq \left(1 - \frac{u(B)}{\sup m}\right)^\ell \quad \text{for all } x \in \mathcal{X}.$$

Taking sup over $B$ concludes the proof. $\square$

By Proposition 1, if $m$ is bounded and $u > 0$ then $J$ is *uniformly ergodic*, in the sense that $\|J^\ell(x,\cdot) - P\| \leq q\rho^\ell$, $x \in \mathcal{X}$, for some constants $q$ and $\rho \in (0,1)$ (here, $q = 1$ and $\rho = 1 - \frac{u}{\sup m}$). To fix ideas, this happens in case $\mathcal{X}$ is compact, $\Theta$ a Polish space, $m$ bounded, and $f$ strictly positive and continuous. An example of DKS falls in this class.

EXAMPLE 4.1.1 (BETA/BINOMIAL). Let $\pi$ be uniform, so that $m(x) = 1/(n+1)$ for all $x \in \mathcal{X} = \{0,1,\ldots,n\}$. Taking sup over those $B$ of the form $B = [\delta, 1-\delta]$, $0 < \delta < 1/2$, yields $u \geq \frac{1}{n+1}(\frac{n}{2(n+1)})^n$. Thus, Proposition 1 gives $\|J^\ell(x,\cdot) - P\| \leq \rho^\ell$ for all $x$ with

$$\rho = 1 - \left(\frac{n}{2(n+1)}\right)^n.$$

Instead, DKS obtain bounds for $x = n$ only; see Proposition 1.1. More precisely,

$$\frac{1}{2}\beta_1^\ell \leq \|J^\ell(n,\cdot) - P\| \leq \frac{\beta_1^{-1/2}}{1 - \beta_1^{2\ell-1}} \beta_1^\ell$$
$$\text{where } \beta_1 = 1 - \frac{2}{n+2}.$$

Hence, DKS's estimate of the convergence rate, that is $\beta_1$, is (much) better than our $\rho$ for large values of $n$.

## 4. GEOMETRIC ERGODICITY

We first recall a general result on Markov chains.

THEOREM 2 [Rosenthal (1995)]. *Let $K$ be an ergodic Markov kernel on $(S, \mathcal{B})$ with stationary distribution $P$. Suppose*

(4) $$Kg(s) \leq \alpha + \beta g(s), \quad s \in S,$$

*for some measurable function $g: S \to \mathbb{R}^+$ and constants $\alpha$ and $\beta \in (0,1)$. Fix $d > 2\alpha/(1-\beta)$, define $D = \{s \in S : g(s) \leq d\}$ and suppose also that*

(5) $$K(s, \cdot) \geq \epsilon Q(\cdot), \quad s \in D,$$

*for some $\epsilon > 0$ and probability $Q$ on $\mathcal{B}$. Then, for all $r \in (0,1)$ and $s \in S$,*

$$\|K^\ell(s, \cdot) - P\| \leq (1-\epsilon)^{r\ell} + t^\ell \left(1 + \frac{\alpha}{1-\beta} + g(s)\right)$$

$$\text{where } t = \frac{(1 + 2\alpha + 2\beta d)^r (1 + 2\alpha + \beta d)^{1-r}}{(1+d)^{1-r}}.$$

In a Gibbs sampling framework, Theorem 2 turns into:

PROPOSITION 3. *Suppose condition (2) holds and*

(6) $$J\phi(x) \leq \alpha + \beta \phi(x), \quad x \in \mathcal{X},$$

*for some measurable function $\phi: \mathcal{X} \to \mathbb{R}^+$ and constants $\alpha$ and $\beta \in (0,1)$. Fix $d > 2\alpha/(1-\beta)$, define $A = \{x \in \mathcal{X} : \phi(x) \leq d\}$ and suppose also that*

(7) $$\sup_A m < \infty \quad \text{and} \quad \inf_{A \times B} f > 0$$

$$\text{for some } B \in \mathcal{G} \text{ with } P(A \times B) > 0.$$

*Then, for all $r \in (0,1)$ and $x \in \mathcal{X}$,*

$$\|J^\ell(x, \cdot) - P\|$$
$$\leq (1-\epsilon)^{r\ell} + t^\ell \left(1 + \frac{\alpha}{1-\beta} + \phi(x)\right)$$

*with $t$ as in Theorem 2 and $\epsilon = \frac{\pi(B) \inf_{A \times B} f}{\sup_A m}$.*

PROOF. By (2), $J$ is ergodic. By (6), condition (4) holds with $K = J$ and $g(x, \theta) = \phi(x)$. By (7), there is $B \in \mathcal{G}$ with $\inf_{A \times B} f > 0$ and $\pi(B) \geq P(A \times B) > 0$. Since $\sup_A m < \infty$, the same calculation as in the proof of Proposition 1 yields

$$J(x, C) \geq \frac{\pi(B) \inf_{A \times B} f}{\sup_A m} P(C \mid T \in B)$$

for all $x \in A$ and $C \in \mathcal{F} \otimes \mathcal{G}$.

Thus, (5) holds with $\epsilon = \frac{\pi(B) \inf_{A \times B} f}{\sup_A m}$ and $Q(\cdot) = P(\cdot \mid T \in B)$. An application of Theorem 2 concludes the proof. $\square$

Proposition 3 applies to most DKS's examples providing reasonable estimates. Note that: (i) Condition (2) holds (in fact, (3) holds) in such examples. (ii) If (7) holds for all $d$, then $t$ can be made arbitrarily close to $\beta$ for suitable $r, d$. There is a trade-off, however, since the choice of $r, d$ affects $(1-\epsilon)^{r\ell}$. (iii) Letting $\psi = 1 + \alpha/(1-\beta) + \phi$, one has

$$t^\ell \psi(x) \leq e^{-c} \quad \text{whenever } \ell \geq \{c + \log \psi(x)\}/|\log t|$$

for all $x \in \mathcal{X}$ and $c > 0$. This can serve to estimate the impact of the initial state $x$. It is roughly of the same order of some DKS's estimates.

EXAMPLE 4.2.1 (POISSON/GAMMA). Let $\pi$ be standard exponential, so that $m(x) = 2^{-x-1}$ for $x \in \mathcal{X} = \{0, 1, \ldots\}$. We take $\phi(x) = x$. In that case, the set $A = \{\phi \leq d\}$ meets condition (7) for all $d > 0$. As to (6), it suffices noting that

$$J\phi(x) = \frac{1}{m(x)} \int \int a f(a,b) \mu(da) f(x,b) \pi(db)$$
$$= 2^{x+1} \int_0^\infty b f(x,b) e^{-b} \, db$$
$$= \frac{2^{x+1}}{x!} \int_0^\infty b^{x+1} e^{-2b} \, db = \frac{x+1}{2}.$$

Hence, Proposition 3 applies with $\alpha = \beta = 1/2$. Now, acting on $r, d$, upper bounds on the convergence rate can be easily obtained. At this stage, using numerical evaluations is convenient.

EXAMPLE 4.3 (GAUSSIAN). Suppose $\sigma^2 + \tau^2 = 1/2$ and $\pi$ is $N(0, \tau^2)$, so that the posterior distribution $\pi(\cdot \mid x)$ is $N(2\tau^2 x, 2\tau^2 \sigma^2)$. We take $\phi(x) = |x|$. Again, $A = \{\phi \leq d\}$ meets (7) for all $d > 0$. Recalling $E|N(0,1)| = \sqrt{2/\pi}$, one obtains

$$J\phi(x) = \int \int |a| f(a,b) \, da \, \pi(db \mid x)$$
$$\leq \int \{|b| + \sigma \sqrt{2/\pi}\} \pi(db \mid x)$$
$$\leq \sigma \sqrt{2/\pi} + \sqrt{2} \sigma \tau \sqrt{2/\pi} + 2\tau^2 |x|$$
$$= \alpha + 2\tau^2 |x|,$$

say. Since $2\tau^2 < 2(\sigma^2 + \tau^2) = 1$, condition (6) holds with $\beta = 2\tau^2$. Again, acting on $r, d$, one gets estimates (even if non optimal) of the convergence rate.

## 5. HIGHER COMPONENT PROBLEMS AND CONCLUDING REMARKS

Apparently, DKS's argument does not apply to the $k$-component Gibbs sampler when $k > 2$. On the other hand, Propositions 1 and 3 can be adapted to any value of $k$. We illustrate this point with regard to Proposition 1 for $k = 3$. To this end, notation needs to be updated. Suppose $(\mathcal{X}, \mathcal{F})$ is the product of two measurable spaces $(\mathcal{X}_1, \mathcal{F}_1)$, $(\mathcal{X}_2, \mathcal{F}_2)$ and $P$ has a density $f$ with respect to $\mu_1 \times \mu_2 \times \pi$, where $\mu_i$ is a $\sigma$-finite measure on $\mathcal{F}_i$, $i = 1, 2$. The marginal densities of the pairs $x = (x_1, x_2)$, $(x_1, \theta)$ and $(x_2, \theta)$ are assumed finite and strictly positive everywhere. Also, $h$ denotes the density of $(x_1, \theta)$. Then, Proposition 1 takes the form:

PROPOSITION 4. *Let $J$ be the Markov kernel of the 3-component Gibbs sampler. If $m$ and $h$ are bounded, then*

$$\|J^\ell(x, \cdot) - P\| \leq \left(1 - \frac{v}{\sup m}\right)^\ell \quad \text{for all } x \in \mathcal{X}$$

*where* $v = \mu_2(\mathcal{X}_2) \sup_{B \in \mathcal{G}} \pi(B) \frac{\{\inf_{x \in \mathcal{X}, \theta \in B} f(x, \theta)\}^2}{\sup_{x_1 \in \mathcal{X}_1, \theta \in B} h(x_1, \theta)}.$

Incidentally, we note that $\mu_2(\mathcal{X}_2) < \infty$ whenever $\inf_{\mathcal{X} \times B} f > 0$ for some $B \in \mathcal{G}$ with $\pi(B) > 0$.

Next, we would like to draw the Authors' attention to an issue that might potentially enlarge the scope of their argument. Consonni and Veronese (2001) introduced the concept of *conditionally reducible natural exponential families*. Basically, they are multivariate natural exponential families whose densities can be expressed as a product of lower dimensional (possibly univariate) conditional exponential families, each being indexed by its own natural parameter. The underlying idea is intimately related to that of a *cut*. Examples include the multinomial and Wishart sampling families. We wonder whether the methods described by the Authors could be applied recursively to conditionally reducible families admitting a factorization in terms of univariate exponential families, such as the multinomial family.

To sum up, DKS's estimates behave excellently, indeed very close to optimum, in those examples for which they are thought. One further merit is that lower bounds are provided as well. On the other hand, Propositions 1 and 3, presented in this discussion, have a broader scope, can be applied for any initial state $x$ (while DKS's bounds are sometimes available for certain $x$ only), but can provide less sharp bounds.